\documentclass[aps,prl,twocolumn,groupedaddress]{revtex4-2}

\usepackage{graphicx}
\usepackage{amsmath}

\begin{document}


\title{Ghost imaging with zero photons}


%
%
%
\author{Meixue Chen}
\thanks{These authors contributed equally to this work.}

\author{Yiqi Song}
\thanks{These authors contributed equally to this work.}

\author{Yu Gu}

\author{Huafan Zhang}

\author{Huaibin Zheng}

\author{Yuchen He}

\author{Hui Chen}
\affiliation{Electronic Materials Research Laboratory, Key Laboratory of the Ministry of Education \& International Center for Dielectric Research, School of Electronic Science and Engineering, Xi’an Jiaotong University, Xi'an 710049, China}

\author{Yu Zhou}
\affiliation{MOE Key Laboratory for Nonequilibrium Synthesis and Modulation of Condensed Matter, Department of Applied Physics, Xi’an Jiaotong University, Xi’an, 710049, China}

\author{Fuli Li}
\affiliation{MOE Key Laboratory for Nonequilibrium Synthesis and Modulation of Condensed Matter, Department of Applied Physics, Xi’an Jiaotong University, Xi’an, 710049, China}

\author{Zhuo Xu}
\affiliation{Electronic Materials Research Laboratory, Key Laboratory of the Ministry of Education \& International Center for Dielectric Research, School of Electronic Science and Engineering, Xi’an Jiaotong University, Xi'an 710049, China}

\author{Jianbin Liu}
\email[]{liujianbin@xjtu.edu.cn}
\affiliation{Electronic Materials Research Laboratory, Key Laboratory of the Ministry of Education \& International Center for Dielectric Research, School of Electronic Science and Engineering, Xi’an Jiaotong University, Xi'an 710049, China}


\date{\today}

\begin{abstract}
Ghost imaging was first demonstrated with entangled photon pairs and well-known for its peculiar properties. The signal beam that illuminates the object possesses no spatial resolution, whereas the reference beam, which never interacts with the object, is spatially resolved. Either beam alone cannot retrieve the image, which can only be obtained when the signal and reference beams are correlated.  Here we will report a ghost imaging experiment with even more peculiar properties, in which the image can be reconstructed when no photon interacts with the object or even no photon in neither signal nor reference beam. All the photons interacted with the object are discarded.  Only the time bins with zero photon are employed to retrieve the image, a process referred to as “ghost imaging with zero photons” hereafter.  The reason why ghost image can be retrieved with zero photons is jointly determined by photon-number projection measurement and photon statistics of thermal light. The results are helpful to resolve the debate on the physics of ghost imaging and understand the relation between quantum and classical correlations.
\end{abstract}


\maketitle


After the first experimental realization of ghost imaging with entangled photon pairs \cite{pittman1995optical}, a long-standing debate over its classical or quantum nature continues to attract significant interests in the field \cite{erkmen2010ghost,shih2011physics,shapiro2012physics}. Entanglement was thought as a prerequisite for ghost imaging \cite{abouraddy2001role}, which may be the first strict attempt to answer the question that whether it is possible to realize ghost imaging with classical light discussed in the original ghost imaging paper \cite{pittman1995optical}. One year later, Bennink \textit{et al.} gave an alternative answer to the question and experimentally verified that ghost imaging can be mimicked with two classically correlated laser light beams \cite{bennink2002two}. Even through their imaging is more like a projection \cite{dangelo2003can}, Bennink \textit{et al.}’s work inspired a series of studies on this topic and it was confirmed that ghost imaging can be realized with classical thermal light by several groups  \cite{gatti2004ghost,cheng2004incoherent,cai2005ghost,valencia2005two}. 

Erkmen \textit{et al.} attributed ghost imaging to classical effect, regardless of classical or nonclassical light was employed \cite{erkmen2010ghost,erkmen2008gaussian}, while Shih \textit{et al.} interpreted it as fundamentally quantum in both cases \cite{shih2011physics,shih2020introduction}. A moderate opinion on this topic is acknowledging that “all optical imaging phenomena are fundamentally quantum mechanical”, while imaging can be interpreted equivalently in both classical and quantum theories is classical \cite{shapiro2012response}, which agrees with Glauber and Sudarshan’s conclusions \cite{glauber1963photon,sudarshan1963equivalence}. There are also other criteria for differentiating classical and nonclassical phenomenon. For instance, Ragy \textit{et al.} employed quantum discord to analyze ghost imaging with thermal light and concluded that quantum correlation dominates the imaging in low-illumination case \cite{ragy2012nature}. We will confine our discussions to the boundary between classical and nonclassical phenomena defined by the criteria established by Glauber and Sudarshan \cite{glauber1963photon,sudarshan1963equivalence}.

Although there is no final consensus on the physics of ghost imaging, it has found vast applications due to its advantages compared to traditional imaging \cite{padgett2017introduction,shih2024ghost}. Ghost imaging with thermal light can be implemented without lens, which makes it perfect for imaging with THz \cite{olivieri2020hyperspectral,chen2020ghost}, X-Ray \cite{yu2016fourier,pelliccia2016experimental,zhang2018tabletop}, and matter wave \cite{khakimov2016ghost,li2018electron}, in which lens is difficult or even impossible to fabricate. Ghost imaging also inspired some interesting imaging methods due to its unique non-local imaging scheme \cite{altmann2018quantum}. Traditional imaging employs lens to establish point-to-point (spot) correlation between the object and image planes via the first-order interference of light \cite{born2021principles}. The point-to-point (spot) correlation in ghost imaging is formed by the second-order interference of light or two-photon interference \cite{shih2020introduction}. It indicates that as long as there is a point-to-point (spot) correlation between two different planes, the correlation can be employed to form an image.  Lemos \textit{et al.} observed a quantum image with undetected photons based on two-photon interference of entangled photon pairs \cite{lemos2014quantum}. Image can be formed with one \cite{kirmani2014first} or less than one \cite{morris2015imaging} detected photon per pixel in computational imaging. In this paper, we will report a more interesting ghost imaging experiment with thermal light, in which image can be reconstructed via the time bins with zero photon by taking single-photon detectors as photon-number-resolving detectors (PNRDs) to perform photon-number projection measurement. Hereafter, the time bins with zero photon will be referred to as \lq\lq{}zero photons\rq\rq{} for brevity.

\begin{figure}[htb]
\centering
\includegraphics[width=75mm]{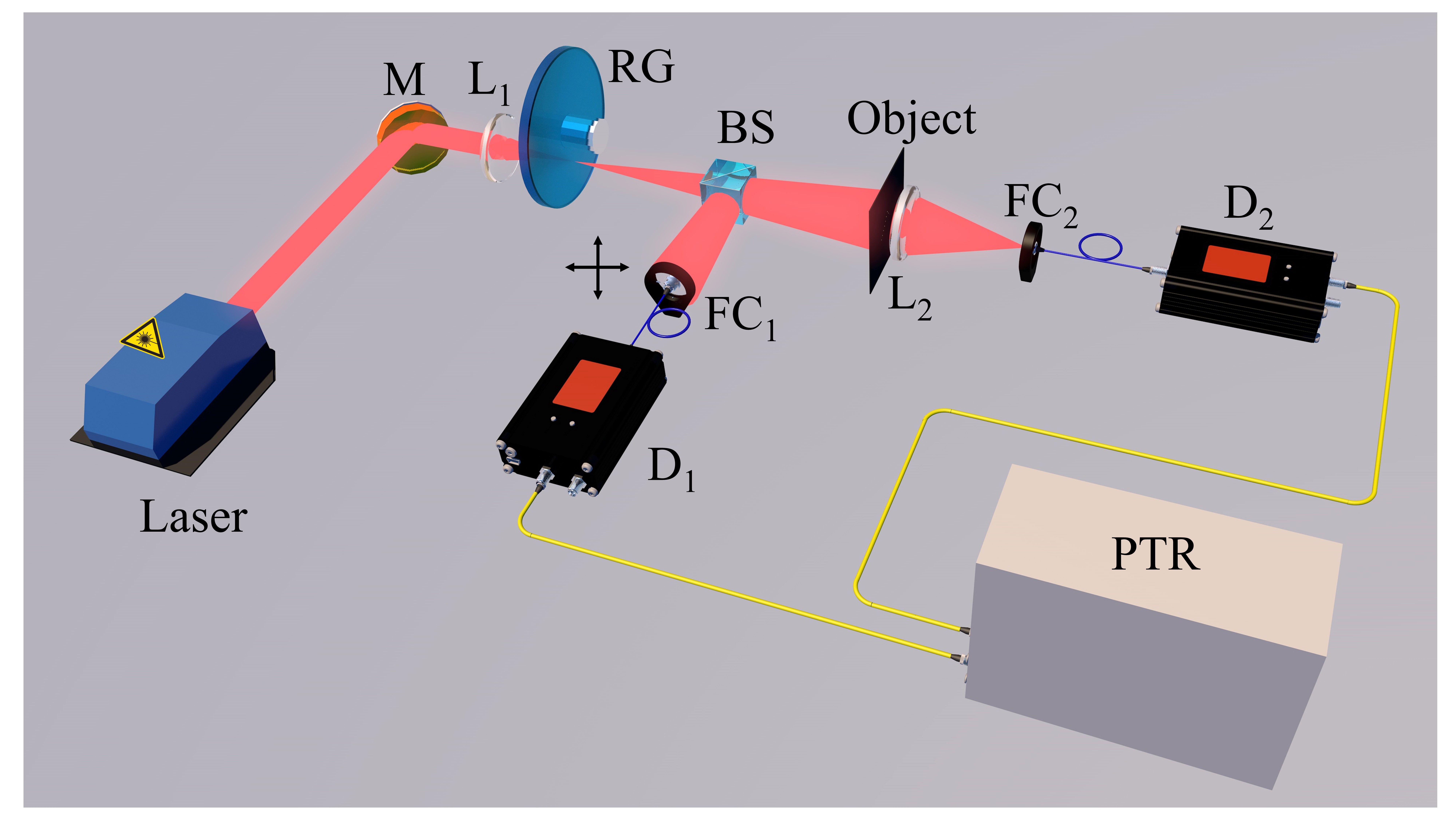}
\caption{ Experimental setup for ghost imaging with zero photons. Laser: single-mode continuous-wave laser. M: mirror. L$_1$ and L$_2$: lens. RG: rotating groundglass. BS: 1:1 non-polarizing beam splitter. Object: transparent object for imaging. FC$_1$ and FC$_2$: fiber couplers. D$_1$ and D$_2$: single-photon detectors. PTR: photon detection time series recording system.}\label{1-scheme}
\end{figure}

The experimental setup for ghost imaging with zero photons is shown in Fig. \ref{1-scheme}, which is similar as traditional setup of ghost imaging with thermal light. The only difference is that two-photon coincidence counts are measured in traditional ghost imaging, while photon detection time series are recorded in our experiment. Pseudothermal light is employed as the light source, which is generated by incident single-mode continuous-wave laser light onto a rotating ground glass (RG). Pseudothermal light is split into two identical parts by a 1:1 non-polarizing beam splitter (BS). One beam passing through the object is collected by a buck detector, which consists of a collecting lens (L$_2$), a fiber coupler (FC$_2$), and a single-photon detector (D$_2$). The reference beam is collected by a scannable fiber coupler (FC$_1$) linked to a single-photon detector (D$_1$). The time series of the photon detection events from D$_1$ and D$_2$ are sent into a photon detection time series recording system (PTR) and recorded independently with the same time clock for later calculations.

\begin{figure}[htb]
\centering
\includegraphics[width=85mm]{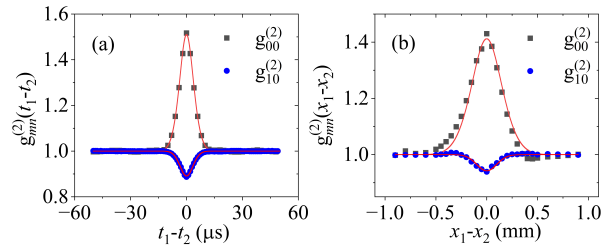}
\caption{Temporal (a) and spatial (b) correlation of zero photons. The black squares and blue circles are experimental results. The red lines are theoretical fittings. $g_{mn}^{(2)}$ is the normalized second-order correlation function of D$_1$ detecting $m$ photons and D$_2$ detecting $n$ photons ($m$, $n$ = 0, 1, 2...).}\label{2-HBT}
\end{figure}

Figure \ref{2-HBT} shows the measured normalized correlation functions of zero photons in a Hanbury Brown-Twiss (HBT) interferometer \cite{hanbury1956correlation}, which can be realized by removing the object and collecting lens in the scheme in Fig. \ref{1-scheme}.  Figure 2(a) was measured when D$_1$ and D$_2$ were in the symmetrical positions. Figure 2(b) was measured by fixing the position of D$_2$ and scanning the position of D$_1$ horizontally. The black squares and blue circles are measured results. The red lines are theoretical fittings of the measured data by employing theoretical results in the following. $g_{mn}^{(2)}$ is the normalized second-order correlation function of D$_1$ detecting $m$ photons and D$_2$ detecting $n$ photons ($m$, $n$ = 0, 1, 2...). For instance, $g_{10}^{(2)}$ is the normalized correlation function of D$_1$ detecting one photon and D$_2$ detecting zero photon. Antibunching is observed in Figs. 2(a) and (b) for  $g_{10}^{(2)}$, in which the degree of second-order correlation, $g_{10}^{(2)}(0)$, is less than 1. More interestingly, the correlation even exists by only considering zero photons for D$_1$ and D$_2$. The point-to-spot correlation in Fig. 2(b) can be employed to perform ghost imaging. 

\begin{figure}[htb]
\centering
\includegraphics[width=85mm]{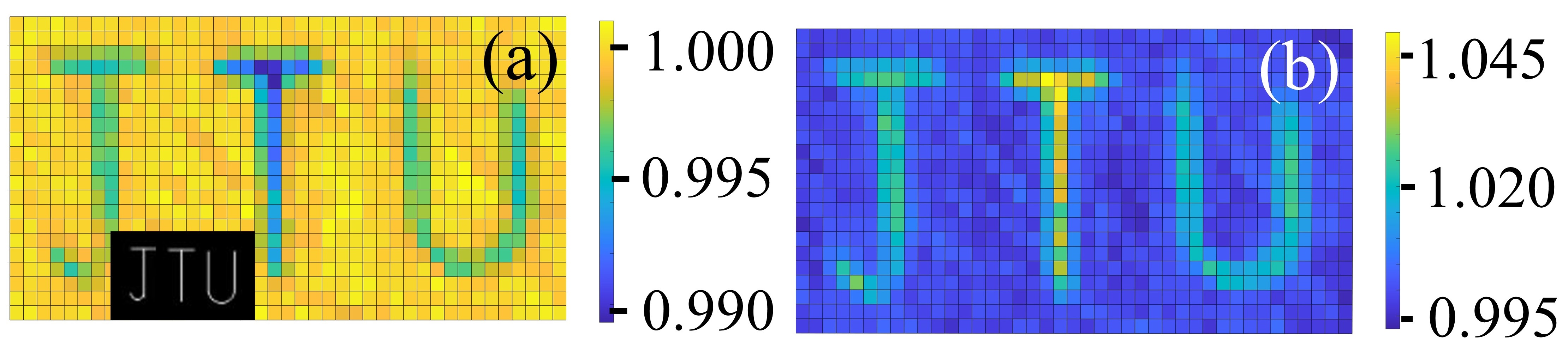}
\caption{(Color online) Ghost images of letters “JTU” with zero photons via  $g_{10}^{(2)}(0)$ (a) and $g_{00}^{(2)}(0)$ (b). The subfigure in (a) is the object. The linewidth of the letters is 0.1 mm.}\label{3-letters}
\end{figure}

Figure \ref{3-letters} shows the reconstructed images of the object consisting with three letters, “JTU”, which are the first letters of our institution, JiaoTong University. The subfigure in Fig. \ref{3-letters}(a) is the object and the line width of the letters is 0.1 mm. The length of horizontal and vertical lines of the letter, “T”, is 1.5 and 3 mm, respectively, which can be served as scale bar. A negative ghost image is observed in Fig. \ref{3-letters}(a) via $g_{10}^{(2)}(0)$, in which all the photons interacted with the object are discarded. A positive ghost image is observed in Fig. \ref{3-letters}(b) via $g_{00}^{(2)}(0)$, in which all the photons in both signal and reference beams are discarded.

The reason why the image can be reconstructed with zero photons can be understood intuitively as follows. There is two-photon bunching for photons in thermal light, which indicates that photons tend to arrive at the detectors in bunches when detectors are at symmetrical positions. There should be positive correlation between zero photons and negative correlation between one photon and zero photon detection events. However, theoretical analysis shows that the physics of ghost imaging with zero photons is not such simple. Negative correlation observed via $g^{(2)}_{10}$ can be tuned into positive correlation by increasing the average number of detected photons. 

It is complicate to calculate the correlation function of D$_1$ detecting $m$ photons and D$_2$ detecting $n$ photons directly \cite{dawkins2024quantum}. With the help of moment generating function \cite{mandel1995optical}, the probability function of D$_1$ detecting $m$ photons and D$_2$ detecting $n$ photons in Fig. \ref{1-scheme} can be obtained via probability generating function \cite{chattamvelli2019generating}. The joint moment generating function in Fig. \ref{1-scheme} is  \cite{mandel1995optical}
\begin{eqnarray}\label{moment}
&&M(u,v;\vec{r}_1,t_1;\vec{r}_2,t_2)\nonumber\\
&\equiv &  \langle e^{u m + v n} \rangle \\
&=& \int_0^\infty dI_1 dI_2\, p(I_1,I_2;\vec{r}_1,t_1;\vec{r}_2,t_2) e^{I_1 T (e^u-1) + I_2 T (e^v-1)} \nonumber,
\end{eqnarray}
where $(\vec{r}_1,t_1)$ and $(\vec{r}_2,t_2)$ are the space-time coordinates of photon detection events at D$_1$ and D$_2$, respectively,  $\langle \cdot \rangle $ is ensemble average, $p(I_1,I_2;\vec{r}_1,t_1;\vec{r}_2,t_2)$ is the joint probability density function, $I_1$ and $I_2$ are the light intensities detected by D$_1$ and D$_2$, respectively, $I_1T$ and $I_2T$ are the average number of detected photons of D$_1$ and D$_2$  in a time bin of width, $T$, respectively. For Gaussian random fields, the joint probability density function is \cite{mandel1995optical}
\begin{eqnarray}\label{pdf}
&&p(I_1, I_2; \vec{r}_1,t_1;\vec{r}_2,t_2) \\
&=& \frac{1}{\langle I \rangle^2 (1-\mu)} \exp\left[ -\frac{I_1+I_2}{\langle I \rangle (1-\mu)} \right] I_0\left( \frac{2\sqrt{\mu I_1 I_2}}{\langle I \rangle (1-\mu)} \right),\nonumber
\end{eqnarray}
where $\langle I \rangle = \langle I_1 \rangle=\langle I_2 \rangle$ is the average intensity detected by D$_1$ and D$_2$, $\mu=|g^{(1)}(\vec{r}_1,t_1;\vec{r}_2,t_2)|^2$, $g^{(1)}(\vec{r}_1,t_1;\vec{r}_2,t_2)$ is the normalized correlation function of light field at $(\vec{r}_1,t_1)$ and $(\vec{r}_2,t_2)$, $I_0\left( \frac{2\sqrt{\mu I_1 I_2}}{\langle I \rangle (1-\mu)} \right)$ is the modified Bessel function of the first kind of order zero. 

Substituting Eq. (\ref{pdf}) into Eq. (\ref{moment}), the moment generating function for thermal light in Fig. \ref{1-scheme} can be simplified as
\begin{eqnarray}\label{moment-final}
&&M(u,v; \vec{r}_1,t_1;\vec{r}_2,t_2)\\
&=&\frac{1}{1 - \bar{n}(e^u-1) - \bar{n}(e^v-1) + (1-\mu) \bar{n}^2 (e^u-1)(e^v-1)},  \nonumber
\end{eqnarray}
where $\bar{n}=\langle I \rangle T$ is the average number of detected photons of D$_1$ or D$_2$. Letting $x=e^u$ and $y=e^v$ in Eq. (\ref{moment-final}), the joint probability generating function of thermal light in Fig. \ref{1-scheme} equals
\begin{eqnarray}\label{probability}
&&M(x,y; \vec{r}_1,t_1;\vec{r}_2,t_2)\\
&=&\frac{1}{1 - \bar{n}(x-1) - \bar{n}(y-1) + (1-\mu)\bar{n}^2 (x-1)(y-1)}.  \nonumber
\end{eqnarray}
The probability function of D$_1$ detecting $m$ photons and D$_2$ detecting $n$ photons is the Taylor expansion coefficient of $M(x,y; \vec{r}_1,t_1;\vec{r}_2,t_2)$ when $x=y=0$ \cite{chattamvelli2019generating},
\begin{eqnarray}\label{probability-final}
&&P_{mn}(\vec{r}_1,t_1;\vec{r}_2,t_2) \nonumber\\
&=& \left. \frac{1}{m! \, n!} \frac{\partial^{m+n} M(x, y; \vec{r}_1,t_1;\vec{r}_2,t_2)}{\partial x^m \partial y^n} \right|_{x=0, y=0}.
\end{eqnarray}
It is difficult to obtain the expression for general value of $m$ and $n$. However, the process is straightforward to obtain $P_{mn}(\vec{r}_1,t_1;\vec{r}_2,t_2)$ when $m$ and $n$ are small or in some special cases. For instance, when $m=0$ (or $n=0$), the probability function has a simple form,
\begin{eqnarray}\label{gm0}
P_{m0}(\vec{r}_1,t_1;\vec{r}_2,t_2)= \frac{\bigl[\bar{n} + (1-\mu)\bar{n}^2\bigr]^m}{\bigl[1 + 2\bar{n} + (1-\mu)\bar{n}^2\bigr]^{m+1}}.
\end{eqnarray}

In order to calculate the normalized second-order correlation function, $g^{(2)}_{m0}(\vec{r}_1,t_1;\vec{r}_2,t_2)$, the probabilities of D$_1$ detecting $m$ photons and D$_2$ detecting zero photon should be calculated separately. The probability can be obtained via photon statistics of thermal light \cite{loudon2000quantum}
\begin{equation}\label{pm}
P_m = \frac{\bar{n}^m}{(1+\bar{n})^{m+1}}, \quad m=0,1,2,\dots
\end{equation}
where $\bar{n}$ is the average number of photons detected by individual photon detector. With Eqs. (\ref{gm0}) and (\ref{pm}), the normalized second-order correlation function of D$_1$ detecting $m$ photon and D$_2$ detecting zero photon equals
\begin{eqnarray}\label{g2-m0}
g^{(2)}_{m0}(\vec{r}_1,t_1;\vec{r}_2,t_2)&\equiv&\frac{P_{m0}(\vec{r}_1,t_1;\vec{r}_2,t_2)}{P_m(\vec{r}_1,t_1)P_0(\vec{r}_2,t_2)} \\
&=&\frac{(1+\bar{n})^{m+2}\bigl[1 + (1-\mu)\bar{n}\bigr]^m}{ \bigl[1 + 2\bar{n} + (1-\mu)\bar{n}^2\bigr]^{m+1}}.\nonumber
\end{eqnarray}

Once the correlation function, $g^{(1)}(\vec{r}_1,t_1;\vec{r}_2,t_2)$, is given, the normalized correlation function $g^{(2)}_{m0}(\vec{r}_1,t_1;\vec{r}_2,t_2)$ can be obtained via Eq. (\ref{g2-m0}). For example, let us consider the case when the two detectors are placed at symmetrical positions and the spectrum of thermal light is Gaussian, i.e., $S(\omega) \propto \exp[-(\omega-\omega_0)^2/(2\sigma^2)]$, where $\omega_0$ is the central angular frequency and $\sigma$ is the standard deviation. Under these conditions, the normalized second-order temporal correlation function of D$_1$ detecting one photon and D$_2$ detecting zero photon is
\begin{eqnarray}\label{g10}
g^{(2)}_{10}(\tau) = \frac{ (1+\bar{n})^3 \bigl[ 1 + \bigl(1 - e^{-\sigma^2 \tau^2}\bigr)\bar{n}\bigl]}{\Bigl[1 + 2\bar{n} + \bigl(1 - e^{-\sigma^2 \tau^2}\bigr)\bar{n}^2\Bigr]^2},
\end{eqnarray}
where $\tau=t_1-t_2$ is the time difference between these two detection events, $\mu = |g^{(1)}(\tau)|^2 = e^{-\sigma^2 \tau^2}$ has been employed to simplify the expression \cite{loudon2000quantum}. Similar method can be employed to calculate other correlation functions such as $g^{(2)}_{00}(\tau)$, $g^{(2)}_{10}(\vec{r}_1,\vec{r}_2)$, and $g^{(2)}_{00}(\vec{r}_1,\vec{r}_2)$.

\begin{figure}[htb]
\centering
\includegraphics[width=65mm]{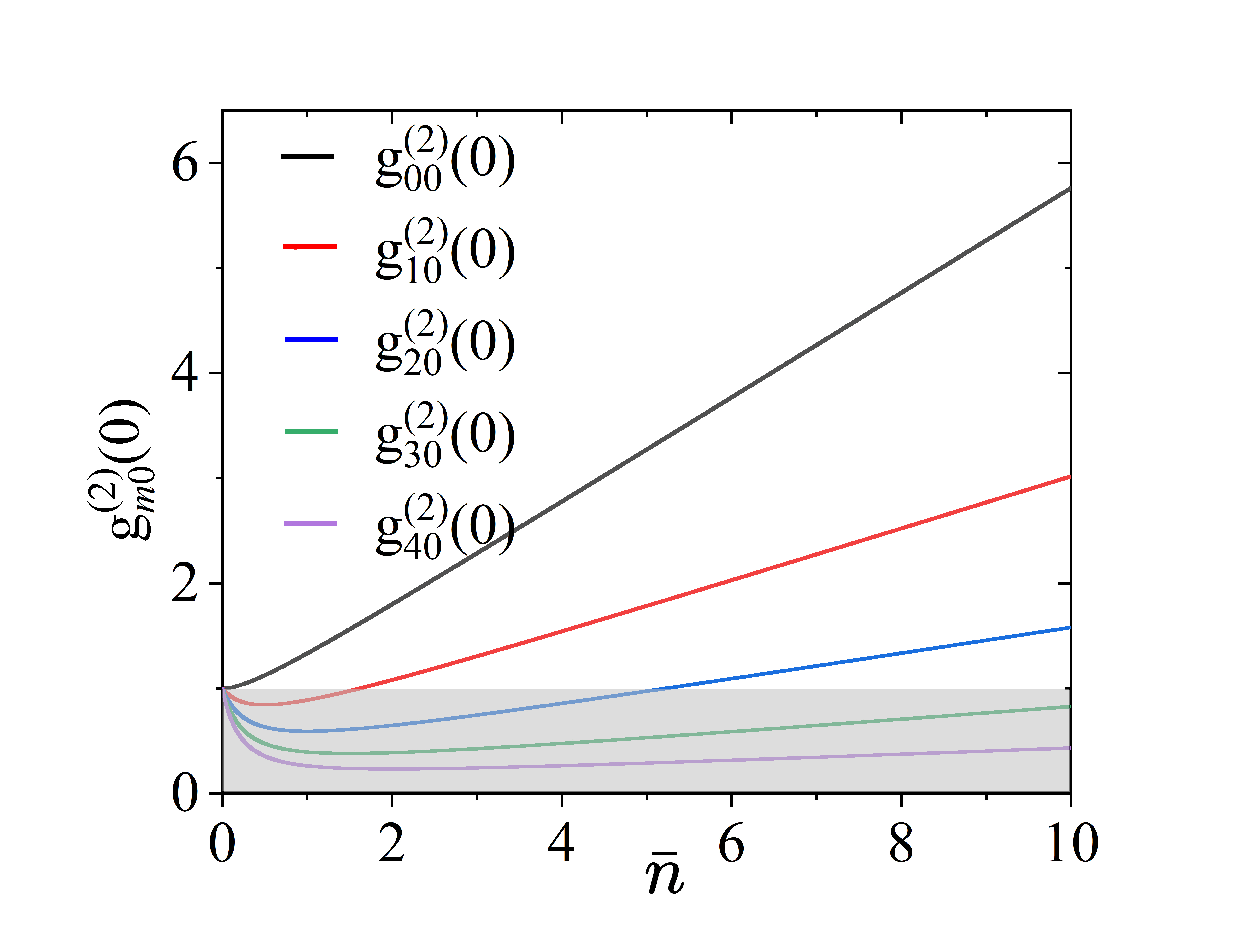}
\caption{Numerical simulations of $g^{(2)}_{m0}(0)$ vs. the average number of detected photons for $m=0$, 1, 2, 3 and 4. The correlation in the gray regime is less than 1 and antibunching can be observed in this regime.}\label{4-average}
\end{figure}

Figure \ref{4-average} shows the degree of the second-order correlation, $g^{(2)}_{m0}(0)$, is not only dependent on the value of $m$, but also on the average number of detected photons, $\bar{n}$. The correlation in the gray regime is less than 1, indicating that antibunching can be observed in this regime. Antibunching can be changed into bunching by increasing the average number of detected photons. $g^{(2)}_{00}(0)$ can not be less than 1, which means that zero photons are always bunched. It is interesting to notice that $g^{(2)}_{00}(0)$ can be much larger than 2, indicating that zero photons can be more bunched than photons in thermal light. 

\begin{figure}[htb]
\centering
\includegraphics[width=85mm]{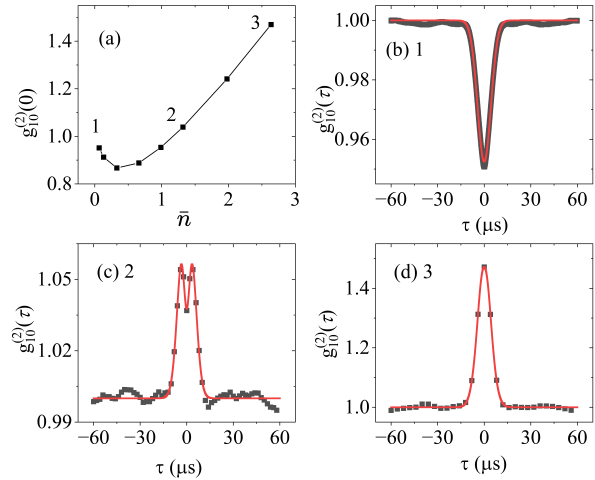}
\caption{Measured $g^{(2)}_{10}(0)$ for thermal light with different widths of time bin (a) and  (b) - (d) are the corresponding temporal correlation functions at three different cases marked 1, 2 and 3 in (a), respectively.  The black squares are experimental data and the red curves are theoretical fittings by employing Eq. (\ref{g10}). The black line in (a) serves as a guide to the eye connecting these dots.}\label{5-g10}
\end{figure}

In order to verify the numerical simulations in Fig. \ref{4-average}, Fig. \ref{5-g10}(a) shows the measured dependence of $g^{(2)}_{10}(0)$ on $\bar{n}$ by varying the width of time bin. Antibunching is observed when $\bar{n}$ is small. When $\bar{n}$ equals 0.5,  $g^{(2)}_{10}(0)$ reaches its minimum. Then antibunching is gradually changed into bunching as $\bar{n}$ continuous to increase.  Figures \ref{5-g10}(b) - (d) are the normalized temporal correlations with different average number of photons labeled with 1, 2 and 3 in Fig. \ref{5-g10}(a), respectively. The black squares in Fig. \ref{5-g10} are experimental results. The black line in Fig. \ref{5-g10}(a) is served as the guidance of eye and the red lines in Figs. \ref{5-g10}(b) - (d) are theoretical fittings of the data by employing Eq. (\ref{g10}). During the fitting, a coefficient (value between 0 and 1) is added in front of  $e^{-\sigma^2 \tau^2}$ to  control the height of the peak, which is the same as the measured degree of second-order coherence of thermal light can not reach 2. It is interesting to notice that $g^{(2)}_{10}(0)$ is less than $g^{(2)}_{10}(\tau)$ when $\tau$ is small in Fig. \ref{5-g10}(c). However, bunching is observed in this case since the value of $g^{(2)}_{10}(0)$ exceeds 1.
 
The reason why antibunching can be changed into bunching in $g^{(2)}_{10}$ as $\bar{n}$ increases is as follows. When $\bar{n}$ is large, the probability for the time bin with one or zero photon is extremely small, which is mainly concentrated in the tail part of the geometrical distribution of thermal light and is named as \lq\lq{}long tail effect\rq\rq{} for short. The probability of D$_1$ detecting one photon and D$_2$ detecting zero photon is a low-probability event. $P(1,0)$ equals $ \frac{\bar{n}}{(1+2\bar{n})^2} \sim \frac{1}{4\bar{n}}$ for $\bar{n} \gg 1$. The normalized correlation function,  $g^{(2)}_{10}$, is defined as $\frac{P(1,0)}{P(1)P(0)}$, where $P(1)P(0)=\frac{\bar{n}}{(1+\bar{n})^2} \times \frac{1}{1+\bar{n}} \sim \frac{1}{\bar{n}^2}$ is the probability of D$_1$ detecting one photon and D$_2$ detecting zero photon when these two events are independent. The normalized correlation function, $g^{(2)}_{10}(0)$, is proportional to $\bar{n}/4$, which indicates that these two detection events are bunched when $\bar{n}$ is large. 

\begin{figure}[htb]
\centering
\includegraphics[width=85mm]{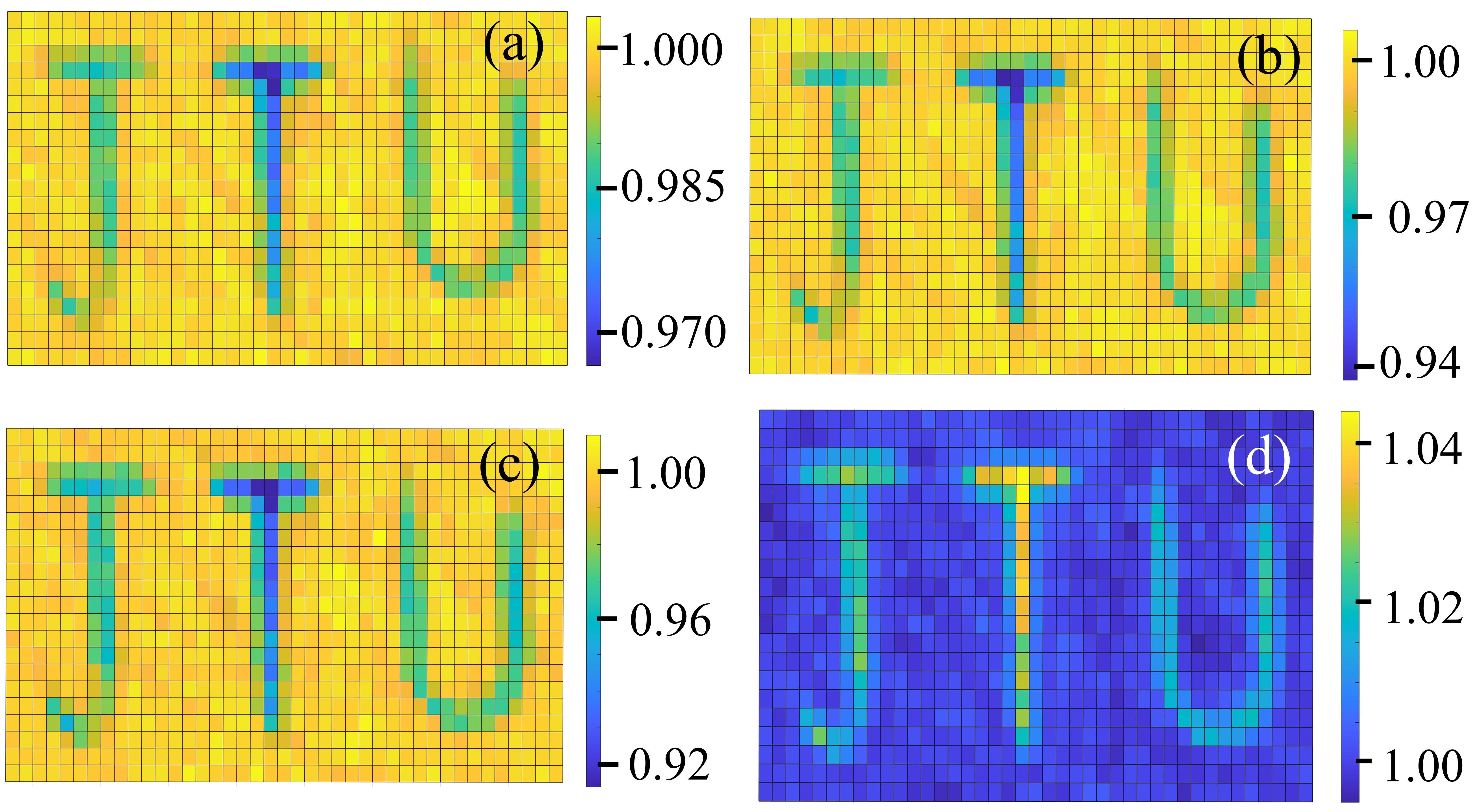}
\caption{(Color online)  Ghost images with  $g_{20}^{(2)}(0)$ (a),  $g_{30}^{(2)}(0)$ (b), and  $g_{40}^{(2)}(0)$  (c) and with traditional ghost imaging algorithm, $g^{(2)}(0)$ (d).}\label{6-letters}
\end{figure}

As predicted in Fig. \ref{4-average}, the visibility of the reconstructed images in Fig. \ref{3-letters}(a) via $g^{(2)}_{10}(0)$ can be increased by employing $g^{(2)}_{m0}(0)$ when $m$ is larger than 1. Figures 6(a) – (c) are the reconstructed images by treating single-photon detectors as PNRDs with resolution larger than 1 with time-multiplexing technique \cite{achilles2004photon,cao2024quantum,mostafavi2025multiphoton}.  The length of time bin is 1 $\mu$s for calculation. All the images in Fig. \ref{6-letters} are calculated from the same data as the one used for the images in Fig. \ref{3-letters}. The visibility increases from 0.0031 to 0.0271 for the images reconstructed via $g_{10}^{(2)}(0)$ and  $g_{40}^{(2)}(0)$, of which the visibility increases nearly 8 times. The peak signal-to-noise ratio of these images does not increase, which are 10.6323, 10.6405, 10.6586, and 10.6747 dB for the ones reconstructed via $g_{10}^{(2)}(0)$  to  $g_{40}^{(2)}(0)$, respectively. For comparison, Fig. \ref{6-letters}(d) shows the reconstructed image with traditional ghost imaging algorithm and the visibility equals 0.0140. 

The classical or nonclassical nature of ghost imaging cannot be solely determined by the classical or nonclassical nature of the employed light. Ghost imaging with classical light can be nonclassical on condition that the detection scheme is nonclassical. Single-photon detector can be either classical or nonclassical based on how the detected information is processed. For example, negative ghost image was reported in several studies \cite{wu2011two,meyers2012positive,cao2024quantum,mostafavi2025multiphoton}. If single-photon detector is  employed to count the number of detected photons, there is no intrinsic difference between negative ghost imaging with single-photon detectors \cite{meyers2012positive} and intensity detectors \cite{wu2011two}. However, if single-photon detector is used as PNRD to perform photon-number projection measurement, it can be treated as a nonclassical detection scheme \cite{cao2024quantum,mostafavi2025multiphoton}, which is different from intensity detector. On the other hand, even if nonclassical light is employed, the imaging may be classical on condition that only classical correlation is employed in the imaging process. 

Unlike classical mechanics, the detection scheme is crucial for the observed phenomena in quantum mechanics \cite{wheeler2014quantum}. The physics of traditional ghost imaging with thermal light \cite{gatti2004ghost,cheng2004incoherent,cai2005ghost,valencia2005two} is the same as that of two-photon bunching of thermal light \cite{liu2024quantum}.  The original spatial two-photon bunching experiment \cite{hanbury1956correlation} can be seen as a special ghost imaging experiment with thermal light, in which the object is a pinhole.  The debate over its classical or nonclassical nature of two-photon bunching of thermal light has been settled \cite{glauber1963photon,sudarshan1963equivalence}. However, the debate repeated itself when ghost imaging with thermal light was discovered \cite{erkmen2010ghost,shih2011physics,shapiro2012physics}. The conclusions about the physics of two-photon bunching of thermal should be valid for ghost imaging with thermal light. While this work focuses on ghost imaging with zero photons using thermal light, extending our scheme to other light sources presents a natural and promising extension. For instance, exploring ghost imaging with zero photons with entangled photon pairs would be an interesting topic.

In conclusion, we have reported a new type of ghost imaging with thermal light via zero photons, in which all the photons interacted with the object are discarded. By employing the time bins with zero photon detection event, the image can be reconstructed by calculating the correlation of zero photons.  The image can even be reconstructed when all the detected photons  in both the signal and reference beams are discarded. The physics of ghost imaging with zero photons are analysed and it is concluded that ghost imaging with classical light can be nonclassical on condition that nonclassical detection scheme is employed. Single-photon detector working as photon-number-resolving detector is a nonclassical detection scheme. Our results are helpful to resolve the longstanding debate on the physics of ghost imaging and may open up a new research direction in quantum optics. 

\begin{acknowledgments}
This work was supported by the Xi'an Science and Technology Program Project (Grant No. GX2331), Shaanxi Key Research and Development Project (2024CY2-GJHX-89), and National Training Program of lnnovation for Undergraduates.
\end{acknowledgments}

\bibliography{antibunching.bib}

\end{document}